\documentclass[12pt,reqno]{amsart}
\newcommand{\D}{\displaystyle}

\usepackage{mathrsfs}
\usepackage{multicol}
\usepackage{epsfig}
\usepackage{latexsym}
\usepackage{amsmath}
\usepackage{enumerate}
\usepackage{amssymb}
\usepackage{amsfonts}
\usepackage{graphicx}
\usepackage{amscd}
\usepackage{verbatim}
\usepackage{calc}
\begin{document}

\begin{center}
\textbf{Hilbert  Manifold Structure of The Set of Solutions of Constraint Equations For Coupled Einstein and Scalar Fields}
\end{center}
	
\begin{center}
Juhi H. Rai \\
Department of Mathematics, RTM Nagpur University, University Campus, Nagpur-440033, India \\
Mail : juhirai$28@$gmail.com  \\
R.V.Saraykar\\
Department of Mathematics, RTM Nagpur University, University Campus, Nagpur-440033, India\\
Mail : ravindra.saraykar$@$gmail.com\vspace{0.5cm}\\
\end{center}
 \textbf{Abstract}: In this paper, we prove that the set of solutions of constraint equations for coupled Einstein and scalar fields in classical general relativity possesses Hilbert manifold structure. We follow the work of R. Bartnik [2] and use weighted Sobolev spaces and Implicit Function Theorem to prove our results.   
\textbf{2000 MR (Mathematical Reviews) Subject classifications :} 46E35, 58D17, 58J05, 83C05.  \\\
\textbf{Abbreviated Title :} Hilbert Manifold Structure.\\

\begin{enumerate}[1.]
\item \textbf{INTRODUCTION} \vspace{5pt}\\
In classical general relativity, Einstein field equations coupled with massless scalar fields are described by \\
$R_{\mu \nu}-\frac{1}{2}Rg_{\mu \nu}=\chi T_{\mu \nu}$ where \\
  $T_{\mu \nu}=2\beta[2\ ^4\psi_{,\mu}\ ^4\psi_{,\nu}- ^4g_{\mu \nu}(\ ^4\psi_{,p}\ ^4\psi^{,p})]$ \\
Here $^4{\psi}$ denotes a real valued function on $R^4$ such that $\psi=i^{\ast}(^4\psi)$,                      
  $i^{\ast}$ to be explained below.
$T$ is stress energy tensor, $R_{\mu \nu}$ denotes Ricci tensor , $R$ is curvature scalar, $^4g_{\mu \nu}$ is spacetime metric, $\beta$ being positive constant corresponding to the 
choice of units. (choice of units can be made so that $\chi$ may be taken as unity.) Since we are considering massless scalar fields, there is no mass term in $T_{\mu \nu}$. If we include mass term then in certain cases , for example in the case of $\pi^0$ - mesons , energy condition gets violated. (See , for example, Hawking and Ellis $[12]$, pages 95-96).  If we consider a spacetime resulting from evolution of a three dimensional spacelike hypersurface $M$ which is usually taken as three dimensional compact or non-compact Riemannian manifold, then spacetime can be described as $M\times R$ and above field equations can be split into four constraint equations and six evolution equations in terms of three dimensional quantities defined on $M$. This is well-known as ADM formalism (Misner,Thorne and Wheeler $[17]$, Chapter $21$) and splitting uses Gauss-Codazzi equations from differential geometry. These equations are given as follows :\\
Constraint equations:\\
\begin{equation*}
\begin{split}
\Phi_0(g,\psi,\pi, \gamma)=& R(g)\sqrt{g}-(|\pi|^2-1/2(tr_g \pi)^2)/\sqrt{g}\\
& +2\beta[(\gamma^2+A(\psi))]\mu_g=0
\end{split}
\end{equation*}
(This is known as Hamiltonian constraint equation), and
\begin{equation*}
\begin{split}
\Phi_i(g,\psi,\pi, \gamma)=& 2(\nabla^jK_{ij}-\nabla_i (tr_gK))\sqrt{g}+\sigma\psi_{,i}=2g_{ij}\nabla_k\pi^{jk}+\sigma\psi_{,i}\\
=&2g_{ij}\nabla_k\pi^{jk}+4\beta\gamma\mu_g\nabla\psi=0 
\end{split}
\end{equation*}
(This is known as Momentum constraint equation). \\
Here, $\sigma=4\beta\gamma\mu_g$\\ 
$\pi$ is momentum density conjugate to $g$. \\
$\gamma$ is scalar density conjugate to $\psi$. \\
Also $A(\psi)=\psi_{,i}\psi^{'i}=|\nabla\psi|^2$ \\ 
And Evolution equations are as follows: \\
\begin{equation*}
\begin{split}
\partial g/\partial t=& 2N[\pi'-\frac{1}{2}(tr\pi')g]-L_Xg \\
\partial \pi/\partial t=& NS_g(\pi,\pi)-[N\ \text{Ein(g)-Hess}\ N-g\triangle N]\mu_g \\
& +\beta N\gamma^2g\mu_g-\beta N(2\overline{\psi}-gA(\psi))\mu_g-L_X\pi
\end{split}
\end{equation*}

$\partial \psi/\partial t=-\sigma'N/4\beta-L_X\psi$ \ \ \ \ \ \ \ as $\sigma'=\gamma$ \\
$\partial \gamma/\partial t=4\beta N\triangle\psi\mu_g-4\beta (\nabla N.\nabla\psi)\mu_g-L_X\gamma$ \\ 
Here $N$ is the Lapse function and $X$ is the shift vector field, \\
and where $\overline{\psi}=\psi_{,i}\psi_{,j}, \ gA(\psi)=|\nabla\psi|^2g_{ij}$ \\ $ \triangle N = -g^{ij}  N_{|i|j} ,  HessN =N_{|i|j} , \\ Ein(g)=Ric(g) -\frac{1}{2} R(g)g  ,\\ S_g (\pi,\pi)= -2 [\pi'\times\pi' - \frac{1}{2} (tr \pi')\pi'] \mu_g +\frac{1}{2} g^\sharp [\pi'.\pi' - \frac{1}{2} (tr\pi')^2] \mu_g  ,  \\(\pi'\times\pi')^{ij} = (\pi')^{ik} (\pi')^j_k  ,  \pi'.\pi' = (\pi')^{ij} (\pi')_{ij} $\\
For sufficiently smooth metric $g$, if $K$ denotes the second fundamental form, then we have  $\pi'=(K-(tr\ K)g)$ and $\pi=\pi'\otimes \mu_g\,$ where $\mu_g$ is the volume element corresponding to $g$. We consider the Constraint function $\Phi=(\Phi_0,\Phi_i)=\Phi(g,\psi,\pi,\gamma).$ \\
Mathematical aspects of this formalism such as the problem of linearization stability and its relationship with the presence of Killing fields, manifold structure of set of solutions of constraint equations, existence and uniqueness of solutions of constraint equations for vacuum spacetime as well as spacetime with matter fields such as electromagnetic fields, Yang-Mills fields, scalar fields etc. attracted attention of mathematicians and theoretical physicists for more than four decades. These aspects are aptly described in the review articles by Fischer and Marsden $[10]$, York $[21]$, Choquet-Bruhat and York $[7]$ and more recently by Bartnik and Isenberg $[4]$ and also in the recent book by Choquet-Bruhat $[5]$.\\
As far as system of Einstein field equations coupled with scalar fields is concerned, Saraykar and Joshi $[18]$ proved that this system is linearization stable if mean curvature of spacelike hypersurface is constant. Here, spacelike hypersurface was assumed to be compact. Later, using weighted Sobolev spaces developed by Christodoulou and Choquet-Bruhat $[8]$, Saraykar $[19]$ proved that this system is linearization stable even if the spacelike hypersurface is non-compact. Furthermore, Saraykar $[20]$ completed Arms-Fischer-Marsden-Moncrief program $[11,1]$ for this system.\\
Of late, there have been renewed interest in the study of Einstein constraint equations in the sense of studying manifold structure of the set of solutions of these equations. For example, Chrusciel and Delay $[9]$ used weighted Sobolev spaces and weighted Holder spaces along with the Implicit Function Theorem to prove that this set carries a Banach manifold structure, whereas Bartnik $[2]$ used particular weighted Sobolev spaces as Hilbert spaces to prove that this set possesses Hilbert manifold structure.\\
In the present paper, we follow Bartnik's work to prove that the set of solutions of constraint equations for above coupled system carries a Hilbert manifold structure. We assume linearization stability results for this system proved by Saraykar and Joshi $[18]$ and Saraykar $[19]$ as mentioned above.\\
Thus, in Section $2$, we describe appropriate function spaces and their properties. We also describe standard inequalities which are needed for our purpose. In Section $3$, we state our main theorem and prove a number of lemmas which will be required to prove the main theorem. In Section $4$, we give the proof of the main theorem. We conclude the paper with remarks about recent work on non-uniqueness of solutions of Einstein constraint equations and probable future work.\\
\item \textbf{Preliminaries and Notations} \\
In this section we introduce the basic framework and notations used in the paper, and recall some well known expressions related to constraint equations.\\
Let $M$ be a connected, oriented and non-compact, $3$ dimensional manifold, and suppose there is a compact subset $M_0\subset M$ such that there is a diffeomorphism
 $\varphi:M\backslash M_0\rightarrow E_R$, where $E_R\subset R^3$  is an exterior region $E_R =\left\{x\in R^3 :|x|>R\right\}$. We also use the notation $B_R$  for the open ball of radius $R$ centred at $0\in R^3, A_R=B_{2R}\backslash \overline{B_R}$  for the annulus  and $S_R=\partial B_R$ for the sphere of radius $R$. Although we assume $\partial M=\emptyset$ for simplicity,  most  of the results are valid when $\partial M$ is non-empty and consists of a finite collection  of disjoint  compact 
$2$-surfaces. Let $\dot{g}$ be a fixed Riemannian metric on $M$ satisfying $\dot{g}=\varphi^{\ast}(e)$ in $M\backslash M_0$, where $e$ is the natural flat metric on $R^3$. In the terminology of $[3],\ \varphi$ is a structure of infinity on $M$.\\ 
By an asymptotically flat spacetime we mean a Lorentz metric $^4g$ on $R^4$ which, in the Euclidean coordinates on $R^4$ satisfies:\\
$^4g_{\mu \nu}-\eta_{\mu \nu}$ is of class $H_{s,\delta}$. Here $H_{s,\delta}$ denotes an appropriate weighted Sobolev space of class ($s,\delta$) as explained in Christodoulou and Choquet-Bruhat [$8$] and $\eta$ denotes the standard Minkowski metric (diag$(-1,1,1,1)$) on $R^4$. By an asymptotically flat hypersurface $\sum \subset R^4$ of an asymptotically flat space-time $(R^4,\ ^4g)$ we mean a spacelike hypersurface $\sum=i(R^3)$, where $i: R^3\rightarrow R^4$ is a spacelike embedding and such that the induced metric $g\sum$ is given by $ i^{\ast}(\ ^4g)$ (with $\sum$ identified with $R^3$) .  $g$ and the second fundamental form $K$ satisfy (in the Euclidean coordinates on $R^3$) the conditions: $g-e$ is of class $H_{s,\delta}$ and $K$ is of class $H_{s-1,\delta+1}$. Let $S_{s,\delta}$ be the Sobolev space of sections of the tensor bundle of symmetric covariant $2$-tensors on $R^3$ whose components are of class $H_{s,\delta}$. Let $M_{s,\delta}$ denote the set of Riemannian metrics $g$ such that $g-e \in S_{s,\delta}$. Then $M_{s,\delta}$ is an open cone in $S_{s,\delta}+\left\{e\right\}$. Thus the tangent space to $M_{s,\delta}$ at $g$ is $T_g M_{s,\delta}=S_{s,\delta}$. Let $\Lambda_{s,\delta}$ denote one-form densities of class $H_{s,\delta}$.\\
As noted above $^4\psi$ denotes a real valued function on $R^4$ such that $\psi=i^{\ast}( ^4\psi)$ is an element of $H_{s,\delta}(R^3,R)=\mathcal{F}_{s,\delta}$. Let $\hat{\mathcal{F}}_{s,\delta}$ denote the class of scalar densities on $R^3$. We also have $\pi'=(K-(tr\ K)g)$. Then $\pi'\in S_{s-1,\delta+1}$. Denote by $\tilde{S}_{s-1,\delta+1}$ the 2-covariant symmetric tensor densities. Then $\pi=\pi'\otimes \mu_g\in \tilde{S}_{s-1,\delta+1}$, where $\mu_g$ is volume element corresponding to $g$.\\
The above considerations apply to the function spaces which we consider below. \\
Let $r\in C^{\infty}\ (M)$ be some function satisfying $r(x) \ge 1$ for all $x\in M$ and $r(x)=|x|$ for all $x\in M\backslash M_0$. Using $r$ and $\dot{g}$, we define the weighted Lebesgue and Sobolev spaces (cf.$[3]$)$\ L_{\delta}^p, W_{\delta}^{k,p}, 1\le p\le\infty,\ \delta \in R$, as the completions of $C_c^{\infty}(M)$ under the norms :\\
$||u||_{p,\delta}=(\int_M|u|^p r^{-\delta p-3}dv_0)^{1/p}$,\\
$||u||_{k,p,\delta}=\sum\limits_{j=0}^k||\dot{\nabla}^ju||_{p,\delta-j}$ \\
if $p<\infty$, and the appropriate supremum norm if $p=\infty$. Here $C_c^{\infty}(M)$ denotes the space of $C^{\infty}$ functions on $M$ with compact support, and $dv_0, \dot{\nabla}$ are respectively the volume measure and connection determined by the metric $\dot{g}$. The weighted Sobolev space of sections of a bundle $E$ over $M$ is defined similarly and is denoted by $W_{\delta}^{k,p}(E)$. The spaces which will be useful to us are:
$\mathcal{G}=W_{-1/2}^{2,2}(S),\mathcal{K}=W_{-3/2}^{1,2}(\tilde{S}),\mathcal{L}=L_{-1/2}^2(T),\mathcal{L}^{\ast}=L_{-5/2}^2 (T^{\ast}\otimes \Lambda^3)$
where $S=S^2T^{\ast}M$ is the bundle of symmetric bilinear forms on $M$, $\tilde{S}=S^2TM\otimes \Lambda^3T^{\ast}M$ is the bundle of symmetric tensor –valued $3$-forms (densities) on $M$ and $T$ is the bundle of spacelike tangent vectors. \\
Thus, for example, $\mathcal{L}$ is a class of  tangent vector fields on $M,\ \mathcal{L}$ and $\mathcal{L}^{\ast}$ are dual spaces with respect to the natural $\L^2$ pairing. \\
For asymptotically flat metrics, the following Hilbert manifolds modelled on $\mathcal{G}$  are natural domains :\\
$\mathcal{G}^+=\left\{g:g-\dot{g}\in\mathcal{G}, g>0\right\}, \mathcal{G}_{\lambda}^+=\left\{g\in \mathcal{G}^+,\lambda\dot{g}<g<\lambda^{-1}\dot{g}\right\},$\\
$0<\lambda<1$.\\
 \\
For the Einstein field equations coupled with scalar fields, the phase space is the Hilbert Manifold given by $\mathcal{P}$ = $\mathcal{G}^+\times \mathcal{F}_{s,\delta}\times \mathcal{K}\times \hat{\mathcal{F}}_{s-1,\delta+1}$. We assume that $s$ is sufficiently large and $\delta$ chosen appropriately, so that initial data in the phase space satisfies sufficient regularity conditions so as to make well-known existence and uniqueness results applicable to the setting here.\\ 
Range of $\Phi$ is $W_{-5/2}^{0,2}(\mathcal{F}({M})\times W_{-5/2}^{0,2}(\Lambda^1T^{\ast}M)$,  where $\mathcal{F}({M})$ denotes the bundle of 
scalar function densities on $M$ and $\Lambda^1T^{\ast}M$ is the bundle of $1$-forms (densities) on $M$.\\
The functional derivative $D\Phi$ is given formally by 
\begin{equation*}
\begin{split}
D\Phi_0(g,\psi,\pi,\gamma).(\tilde{g},\tilde{\psi},\tilde{\pi},\tilde{\gamma})= & \mu(g)^{-1}\{-\frac{1}{2}\left(\pi.\pi-\frac{1}{2}(tr\pi)^2tr\tilde{g}\right)+2\left(\pi-\frac{1}{2}(tr\pi)g\right)\tilde{\pi}\\
&+2\left(\pi\times\pi-\frac{1}{2}(tr\pi)\pi\right)\tilde{g}-\mu(g)[\delta\delta\tilde{g}+\triangle (tr\tilde{g})-Ein(g)\tilde{g}]\\
& -\beta\gamma^2g\tilde{g}\mu(g)-\beta(2\tilde{\psi}-g|\nabla\psi|^2\tilde{g}\mu (g) +4\beta\gamma\eta\mu(g)+\nabla\psi\nabla\tilde{\psi}\mu(g)
\end{split}
\end{equation*}
\begin{equation*}
\begin{split}
D\Phi_i(g,\psi,\pi,\gamma).(\tilde{g},\tilde{\psi},\tilde{\pi},\tilde{\gamma})=&\\
&-2\tilde{g}\delta_ {g}\pi-2g\delta_ {g}\tilde{\pi}+\pi^{jl}(\tilde{g}_{kj;l}+\tilde{g}_{kl;j}-\tilde{g}_{jl;k})-\gamma\nabla\tilde{\psi}-\tilde{\gamma}\nabla \psi\\
\end{split}
\end{equation*}
The expressions for formal $L^2$ adjoint operator $D\Phi^{\ast}$ are given by \\
$D\Phi_g^{\ast}[N,X]=-2KN-L_Xg = f_1$ ,\\
$D\Phi_{\psi}^{\ast}[N,X]=-\sigma'N/4\beta-L_X\psi = f_2$, \\ 
$D\Phi_{\pi}^{\ast}[N,X]=\sqrt{g}(\nabla^2N-(\triangle gN)g)+(S-E+T)N+L_X\pi = f_3$ and \\
$D\Phi_{\gamma}^{\ast}[N,X]=-4\beta N(\triangle \psi)\mu_g+4\beta (\nabla N.\nabla \psi)\mu_g+L_X\gamma = f_4$. \\
where $\sigma'=\gamma$, and where $T=[2\beta \overline{\psi}-\beta gA(\psi)]\mu_g-\beta\gamma^2g\mu_g ,  \\ E^{ij}=Ric^{ij} -\frac{1}{2} R(g)g^{ij}, and \\S^{ij}= g^{-1}( (tr_g\pi)\pi^{ij} -2\pi^i_k\pi^{jk}+\frac{1}{2}|\pi|^2 g^{ij}-\frac{1}{4} (tr_g\pi)^2g^{ij}$ \\
Further , in abbreviated form we will use S and E for $S^{ij}$ and  $E^{ij}$ respectively.\\ We need certain standard inequalities  which we describe below in the form of the theorem :\\
\textbf{Theorem 2.1}: The following inequalities hold:
\begin{enumerate}[(1)]
	\item If $1\le p\le q\le\infty, \delta_2<\delta_1$ and $u\in L_{\delta_2}^q$, then $||u||_{p,\delta_1}\le C||u||_{q,\delta_2}$ And hence, $L_{\delta_2}^q \subset L_{\delta_1}^p$.
	\item (Holder  Inequality) : If  $u\in L_{\delta_1}^q, v\in L_{\delta_2}^r$ and $\delta=\delta_1+\delta_2$ ,\\
	$1\le p,q,r\le\infty, 1/p=1/q+1/r$, then $||uv||_{p,\delta}\le||u||_{q,\delta_1}||v||_{r,\delta_2}$. 
\item (Interpolation Inequality): For any $\in>0$, there is a $C(\in)$ such that, For all $u\in W_{\delta}^{2,p},\  1\le p\le\infty$, \\ 
$||u||_{1,p,\delta}\le \in ||u||_{2,p,\delta}+C(\in)||u||_{0,p,\delta}$. 
\item (Sobolev Inequality) : If $u\in W_{\delta}^{k,p}$, then \\
$||u||_{np/(n-kp),\delta}\le C||u||_{k,q,\delta}\ \ \ \  $ if $n-kp>0$ and $p\le q\le np/(n-kp)$, \\
$||u||_{\infty, \delta}\le C||u||_{k,p,\delta}\ \ \ \ \ \ \ \ $ if $n-kp<0$ 
\item (Poincare  Inequality) : If $\delta <0$ and $1\le p<\infty$, for any $u\in W_{\delta}^{1,p}$ \\ 
We have, $||u||_{p,\delta}\le C||\dot{\nabla}u||_{p,\delta-1}$ Where $n=3$ is the dimension of $M$.
\item (Morrey's Lemma ) : If $u\in W_{\delta}^{k,p}$ and $0<\alpha\le k-n/p\le1$ then $||u||_{C_{\delta}^{0,\alpha}}\le C||u||_{k,p,\delta}$ where the weighted Holder norm is given by,
\begin{equation*}
\begin{split}
||u||_{C_{\delta}^{0,\alpha}}=& Sup \ x\in M  (r^{-\delta+\alpha}(X).\max\limits_{4|x-y|\le r(x)}|u(x)-u(y)|/|x-y|^{\alpha}) \\
& +\max\limits_{x\in M}(r^{\delta}(x)|u(x)|).
\end{split}
\end{equation*}
\end{enumerate}
\item \textbf{Statement of the Main Theorem and Preliminary Lemmas:} \\
In this section we state our main theorem and prove a number of lemmas which will be required to prove this theorem.\\
Statement of the Main Theorem: \\
\textbf{Theorem 3.1}:  For each $(\in,S_i)\in \mathcal{L}^{\ast}$, the constraint set
$$C(\in, S_i)=\left\{(g,\psi,\pi,\gamma)\in \mathcal{P}:\Phi (g,\psi,\pi,\gamma)=(\in,Si)\right\}$$ is a Hilbert submanifold of $\mathcal{P}$.\\
In particular, the space of solutions of the constraint equations for coupled Einstein and scalar fields, $C=\Phi^{-1}(0,0)=C(0,0)$ is a Hilbert manifold.\\
We begin with the following lemma :\\
\textbf{Lemma 3.2}: \\
Suppose $g\in \mathcal{G}_{\lambda}^+$ for some $\lambda>0, \psi\in \mathcal{F}_{s,\delta} , \pi\in \mathcal{K}$ and $\gamma\in \hat{\mathcal{F}}_{s-1,\delta+1}$. Then there is a constant $c=c(\lambda)$ such that
\begin{equation*}
\begin{split}
& ||\Phi_0(g,\psi,\pi,\gamma)||_{2,-5/2}\le c(1+||(g-\dot{g}),(\psi-e)||^2_{2,2,-1/2}\\
&+||(\pi,\gamma)||^2_{1,2,-3/2})
\end{split}
\end{equation*}
And 
\begin{equation*}
\begin{split}
& ||\Phi_i(g,\psi,\pi,\gamma)||_{2,-5/2}\le c(||\pi||_{1,2,-3/2}(1+||\dot{\nabla}g||_{1,2,-3/2}) \\
& +(||\dot{\nabla}\psi||_{1,2,-3/2}||\gamma||_{1,2,-3/2}))\le c(||\dot{\nabla}\pi||_{2,-5/2}\\
& +||\dot{\nabla}g||_{1,2,-3/2}||\pi||_{1,2,-3/2}+||\dot{\nabla}\psi||_{1,2,-3/2}||\gamma||_{1,2,-3/2}
\end{split}
\end{equation*}
\textbf{Proof:-} we have,\\
$\Phi_0(g,\psi,\pi,\gamma)=R(g)\sqrt{g}-(|\pi|^2-1/2(tr_g\pi)^2/\sqrt{g}+2\beta[(\gamma^2+A(\psi))]\mu g)$
\begin{equation*}
\begin{split}
\Phi_i(g,\psi,\pi,\gamma)&=2(\nabla^jK_{ij}-\nabla_i(tr_gK))\sqrt{g}+\sigma\psi_{,i}=2g_{ij}\nabla_k\pi^{jk}+\sigma\psi_{,i} \\
&=2g_{ij}\nabla_k\pi^{jk}+4\beta\gamma\mu_g\nabla\psi\textsl{}
\end{split}
\end{equation*}
Where $\gamma$ is the quantity conjugate to $\psi$. \\
Also $A(\psi)=\psi_{,i}\psi^{,i}=|\nabla\psi|^2$ .\\
Since $g\in \mathcal{G}_{\lambda}^+,g$ is Holder continuous with holder exponent $\frac{1}{2}$ and we have the global pointwise bounds,
\begin{equation*}
\lambda\dot{g}_{ij}(X)v^iv^j<g_{ij}(X)v^iv^j<\lambda^{-1}g_{ij}(X)v^iv^j\ \text{for\ all}\ x\in M, v\in R^3\tag{1}
\end{equation*}
by weighted Holder and Sobolev inequalities valid for any function or tensor field $u$,
$$||u^2||_{2,-5/2}=||u||_{4,-5/4}||u||_{4,-5/4}$$
For $\delta_2<\delta_1$ we have $||u||_{p,\delta_1}\le c||u||_{q,\delta_2}$ and $L_{\delta_2}^q \subset L_{\delta_1}^p$ , therefore we get $||u||_{4,-5/4}^2\le||u||_{4,-3/2}^2$ \\
Now, by using Weighted Holder inequality and the fact 
$$||u||_{p,\delta}=(\int|u|^p\sigma^{-\delta p-n}dx)^{1/p},\ p<\infty$$
we get $||u||_{4,-3/2}^2=(\int |u|^4\sigma^3\ dx)^{1/4}$\\
here $n=3,\ ||u||_{4,-3/2}^2=[\int|u|^3\sigma^{3/2}dx\int |u|\sigma^{3/2}dx]^{1/2}$ \\
Also using $\int ||f.g||\le ||f||_p||g||_q$ we get,
$\||u||_{4,-3/2}^2\le||u||_{6,-3/2}^{3/2}.||u||_{2,-3/2}^{1/2}$
We now use the following Sobolev inequality,\\ 
If $u\in W_{\delta}^{k,p}$ then, $||u||_{np/n-kp,\delta}\le C||u||_{k,q,\delta}$, if $n-kp>0$ and $p\le q\le np/(n-kp)$. \\
And we get, $||u||_{6,-3/2}^{3/2}.||u||_{2,-3/2}^{1/2}\le C||u||_{1,2,-3/2}^2$ \\
Therefore, 
\begin{equation*}
||u^2||_{2,-5/2}\le C||u||_{1,2,-3/2}^2\tag{2} 
\end{equation*}
Connections corresponding to  $g$ and $\dot{g}$ are related by the difference tensor $A_{ij}^k=\Gamma_{ij}^k-\dot{\Gamma}_{ij}^k$ which may be defined invariably by,
\begin{equation*}
A_{ij}^k= \frac{1}{2} g^{kl}(\dot{\nabla}_ig_{jl}+\dot{\nabla}_jg_{il}-\dot{\nabla}_lg_{ij}) \tag{3}
\end{equation*}
The scalar curvature can be expressed in terms of $\dot{\nabla}$ and $A_{ij}^k$ by,
\begin{equation*}
\begin{split}
R(g) &=g^{jk}Ric(\dot{g})_{jk}+g^{jk}(\dot{\nabla}_iA_{jk}^i-\dot{\nabla}_jA_{ik}^i+A_{jk}^lA_{il}^i-A_{jl}^iA_{ki}^l) \\
&=g^{ik}g^{jl}(\dot{\nabla}_{ij}^2g_{kl}-\dot{\nabla}_{ik}^2g_{jl})+Q(g^{-1},\dot{\nabla}g)+g^{jk}Ric (\dot{g})_{jk}
\end{split}
\end{equation*}
Where $Q(g^{-1},\dot{\nabla} g)$ denotes a sum of terms quadratic in $g^{-1},\dot{\nabla} g$\\
Using (1) ,(2) and (3) we get,
\begin{equation*}
\begin{split}
||R(g)||_{2,-5/2}^2 &\le C\int_M(|\dot{\nabla}^2g|^2+|\dot{\nabla}g|^4+|Ric(\dot{g})|^2)r^2dv_0 \\
&\le C(1+||\dot{\nabla}^2g||_{2,-5/2}^2+||\dot{\nabla}g||_{4,-5/4}^4) \\ 
&\le C(1+||\dot{\nabla}g||_{1,2,-3/2}^4) 
\end{split}
\end{equation*}
As $||\dot{\nabla}^2g||_{2,-5/2}^2=||\dot{\nabla}g||_{4,-5/4}^4$ \\
Using $||u||_{k,p,\delta}=\sum\limits_{j=0}^k||\dot{\nabla}^ju||_{p,\delta-j}$ \\
We have ,
\begin{equation*}
\begin{split}
||g||_{2,2,-1/2}& =\sum\limits_{j=0}^2||\dot{\nabla}^jg||_{2,\delta-j}  \ \ (\dot{\nabla}\dot{g}=0\ \text{as $\dot{g}$ is flat matric}.) \\
& = ||g||_{2,-1/2}+||\dot{\nabla}g||_{2,-3/2}+||\dot{\nabla}^2g||_{2,-5/2} 
\end{split}
\end{equation*}
 Also, we know, $||\dot{\nabla}^2g||_{2,-5/2}\le C||\dot{\nabla}g||_{1,2,-3/2}^2$.\\
Therefore 
\begin{equation*}
\begin{split}
&||\dot{\nabla}g||_{1,2,-3/2}^2\le ||g-\dot{g}||_{2,2,-1/2}^2 \\
\end{split}
\end{equation*}\\

Thus, we have \\
\begin{equation*}
\begin{split}
& ||Riem(g)||_{2,-5/2}\le C(1+||\dot{\nabla}g||_{1,2,-3/2}^2)\le C(1+||g-\dot{g}||_{2,2,-1/2}^2)
\end{split}
\end{equation*}\\
Similar to estimate for $||\dot{\nabla}g||_{1,2,-3/2}^2$, we get, \\
\begin{equation*}
\begin{split}
&||\dot{\nabla}\psi||_{1,2,-3/2}^2\le ||\psi - e||_{2,2,-1/2}^2 \\
\end{split}
\end{equation*}\\
Moreover,  $||\pi^2||_{2,-5/2}\le C ||\pi||_{1,2,-3/2}^2$ , and similar inequality holds true for $||\gamma^2||_{2,-5/2}$. \\

Thus, we get, combining above inequalities, \\
$||\Phi_0(g,\psi,\pi,\gamma)||_{2,-5/2}\le c(1+||(g-\dot{g}),(\psi-e)||_{2,2,-1/2}^2+||(\pi,\gamma)||_{1,2,-3/2}^2)$ \\
Here we have used the standard definition of product norm, namely,\\
\begin {equation*}
 ||(x,y)||^2 = ||x||^2 + ||y||^2\\
 \end {equation*}\\
 Thus, first part of the lemma follows. \\
 For momentum constraint, \\
Since $\nabla_j\pi^{ij}=\dot{\nabla}_j\pi^{ij}+A_{jk}^i\pi^{jk}$ \\
We have, $\Phi_i(g,\psi,\pi,\gamma)=2g_{ij}(\dot{\nabla}_k\pi^{jk}+A_{kl}^j\pi^{kl})+\gamma\dot{\nabla}\psi$ \\
By  Holder inequality we get, 
\begin{equation*}
\begin{split}
& ||\Phi_i(g,\psi,\pi,\gamma)||_{2,-5/2}^2\le C(||\dot{\nabla}\pi||_{2,-5/2}^2+||\dot{\nabla}g||_{1,2,-3/2}^2||\pi||_{1,2,-3/2}^2+ \\
& ||\dot{\nabla}\psi||_{1,2,-3/2}^2||\gamma||_{1,2,-3/2}^2)\\
\end{split}
\end{equation*}\\
\begin{equation*}
\begin{split}
\le C[(||\dot{\nabla}\pi||_{2,-5/2}+||\dot{\nabla}g||_{1,2,-3/2}||\pi||_{1,2,-3/2}+ 
& ||\dot{\nabla}\psi||_{1,2,-3/2}||\gamma||_{1,2,-3/2})]^2 \\
\end{split}
\end{equation*}
By taking square-root, we get second inequality in the Lemma.\\
Lemma (3.2) is thus proved. \\

Thus $\Phi$ is a quadratically bounded map between the Hilbert manifolds $\mathcal{P}=\mathcal{G}^+\times \mathcal{F}_{s,\delta}\times \mathcal{K}\times \hat{\mathcal{F}}_{s-1,\delta+1}$ and $W_{-5/2}^{0,2}(\mathcal{F}({M})\times W_{-5/2}^{0,2}(\Lambda^1T^{\ast}M)$. The polynomial structure of the constraint functionals enables us to show that $\Phi$ is smooth, in the sense that it has infinitely many Frechet derivatives. Thus, we get a corollary as follows:- \\
\textbf{Corollary 3.3:} $\Phi :\mathcal{P}\rightarrow \mathcal{F}({M})\times (\Lambda^1T^{\ast}M)$ is a smooth map of Hilbert manifolds.\\
Proof follows easily from Lemma (3.2).\\
The next step in proving the main theorem is to study the kernel of the adjoint operator $D\Phi (g,\psi,\pi,\gamma)^{\ast}$.\\
The next lamma establishes coercivity of $D\Phi(g,\psi,\pi,\gamma)^{\ast}$ : \\\\
\textbf{Lemma 3.4} \\
If $\xi \in W_{-1/2}^{2,2}$ satisfies $D\Phi^{\ast}(\xi)=(f_1,f_2,f_3,f_4)$ and \\
$(f_1,f_2,f_3)\in W_{-3/2}^{1,2}\times L_{-5/2}^2\times L_{-5/2}^2$,\\
then $||\xi||_{2,2,-1/2}\le c(||f_1||_{1,2,-3/2}+(||f_2,f_3||_{2,-5/2})+C||\xi||_{1,2,0})$.\\
Here $ f_1= D\Phi_g^{\ast}(\xi) , f_2 = D\Phi_\psi^{\ast}(\xi) ,f_3 = D\Phi_\pi^{\ast}(\xi) , f_4 = D\Phi_\gamma^{\ast}(\xi)$, and $C$ depends upon $\dot{g}$, $\lambda$ , and  $\||(g,\psi,\pi,\gamma)||_\mathcal{P}$.  \\
\textbf{Proof:} We follow the proof of Proposition 3.3 of [2], and while doing so, we provide some details in the proof for the sake of better understanding.\\
We have $f_3 = D\Phi_\pi^{\ast}(\xi)$. Rearranging $f_3$ gives , \\
$\nabla^2N = Q - \frac{1}{2} (tr_g Q)g ,$\\
 where $ Q= D\Phi_\pi^{\ast}(\xi)/\sqrt{g} + (E-S-T)N -L_X\pi / \sqrt{g}$ ,\\ and $ D\Phi_{\pi}^{\ast}(\xi)=\sqrt{g}(\nabla^2N-(\triangle gN)g)+(S-E+T)N+L_X\pi$ \\ 
and thus, after simple calculations, we get $ |\nabla^2N|^2 \le \frac{5}{4} |Q|^2 .$
This gives the estimate as follows:\\ 
\begin{equation*}
\begin{split}\\
& ||\dot{\nabla}^2N||_{2,-5/2}\le C(||D\Phi(g,\psi,\pi,\gamma)_1^{\ast}(\xi)||_{2,-5/2})+||N||_{\infty,0} \\
& (||E||_{2,-5/2}+||S||_{2,-5/2}+||T||_{2,-5/2})+||A\dot{\nabla}N||_{2,-5/2} \\
& +||X||_{\infty,0}||\dot{\nabla}\pi||_{2,-5/2}+||\dot{\nabla}X||_{3,-1}||\pi||_{6,-3/2}
\end{split}
\end{equation*}
Here we have used Holder inequality to get \\ 
$||\dot{\nabla}X.\pi||_{2,-5/2}\le ||\dot{\nabla}X||_{3,-1}+||\pi||_{6,-3/2}$ \\
To estimate different terms on the right hand side in the above inequality, we proceed as follows :\\
First we use Sobolev inequality : \\
 $||u||_{\infty,\delta}\le c||u||_{k,p,\delta}$ if $n-kp <0$ \\
and we get , $||u||_{\infty,0}\le c||u||_{1,4,0}$. \\
Then, by using Holder inequality, we have ,\\
 $||u||_{1,4,0}\le c||u||_{1,2,0}^{1/4}||u||_{1,6,0}^{3/4}$ .\\
Again using Sobolev inequality we get , $||u||_{1,6,0}^{3/4}\le||u||_{2,2,0}^{3/4}$ \\
and so $||u||_{1,4,0}\le c||u||_{1,2,0}^{1/4}||u||_{2,2,0}^{3/4}$. \\
Now,by definition of Weighted Lebesgue space , \\$||u||_{k,p,\delta}=\sum\limits_{j=0}^k||\dot{\nabla}^ju||_{p,\delta-j} \\
||u||_{1,2,0}=\sum\limits_{j=0}^1||\dot{\nabla}^ju||_{p,0-j}=||u||_{2,0}+||\dot{\nabla}u||_{2,-1} \\
||u||_{2,2,0}=\sum\limits_{j=0}^2||\dot{\nabla}^ju||_{p,\delta-j}=||u||_{2,0}+||\dot{\nabla}u||_{2,-1}+||\dot{\nabla}^2u||_{2,-2}$\\
And using Young Inequality : $│ab│\le a^p/p + b^q/q$, and taking \\
$a= ||u||_{1,2,0}^{1/4}=[||u||_{2,0}+||\dot{\nabla}u||_{2,-1}]^{1/4}$\\
$ b=||u||_{2,2,0}^{3/4}=[||u||_{2,0}+||\dot{\nabla} u||_{2,-1}+||\dot{\nabla}^2u||_{2,-2}]^{3/4}$,\\
$p=4$,and $q=4/3$\\we get
 $|ab|=||u||_{1,2,0}^{1/4}||u||_{2,2,0}^{3/4}\\
=([||u||_{2,0}+||\dot{\nabla}u||_{2,-1}]^{1/4})+([||u||_{2,0}+||\dot{\nabla}u||_{2,-1}+||\dot{\nabla}^2u||_{2,2}]^{3/4}) \\
\le [||u||_{2,0}+||\dot{\nabla}u||_{2,-1}]+(3/4)||\dot{\nabla}^2u||_{2,-2} \\
\le  \in ||\dot{\nabla}^2u||_{2,-2}+c{\in}^{-3} ||u||_{1,2,0}$ 
for any $\in >0$.\\
 Therefore, $||u||_{\infty,0}\le \in ||\dot{\nabla}^2u||_{2,-2}+c{\in}^{-3}||u||_{1,2,0}$ for any $\in >0$. \\Similarly we can prove that,for any $\delta \in R$,\\ $||u||_{3,\delta}\le\in ||\dot{\nabla}u||_{2,\delta-1}+c{\in}^{-1}||u||_{2,\delta}$.\\
Thus, we get the estimate \\
$||\dot{\nabla}^2N||_{2,-5/2}\le c||f_3||_{2,-5/2}+\in ||\dot{\nabla}^2\xi||_{2-2}+C||\xi||_{1,2,0}$ \\
since $||N||$ and $||X||$ both can be replaced by $||\xi||$ as $\xi=(N,X)$ \
Since, norms of $g,\psi,\pi ,\gamma$ are bounded, norms of $E$, $S$ and $T$ are also bounded.\\
Now, writing \\
$X_{i|jk}=-R_{ijkl}X^l+ X_{(i|j)k}+X_{(i|k)l}–X_{(j|k)l}$  which is valid for any sufficiently smooth $X_i$,\\
and proceeding as in Bartnik [2], we get\\
$$\D||\dot{\nabla}^2X||_{2, -5/2}\D\le c||f_1||_{1,2,-3/2} + \in ||\dot{\nabla}^2\xi||_{2,-2} + c ||\xi||_{1,2,0}$$
Now, $f_2 = D\Phi_\psi^{\ast}(\xi) = -\sigma'N/4\beta- L_X\psi$ \\
Hence, $N=-4\beta\gamma^{-1}[f_2+L_X\psi]$ as $\sigma'=\gamma$ \\   
Therefore, $||\dot{\nabla}^2N||_{2,-5/2}\le c||f_2||_{2,-5/2}+||X.\dot{\nabla}\psi||_{2,-5/2}$ \\
Now consider $||\dot{\nabla}\psi.X||_{2,-5/2}$\\
Using Holder inequality , we get ,
$$||\dot{\nabla}\psi.X||_{2, -5/2}\le ||\dot{\nabla}\psi||_{3, -1}+||X||_{6, -3/2}$$\\
Also, $||u||_{k,p,\delta_1}\le ||u||_{k,p,\delta_2}$ if $\delta_2\le\delta_1$. \\
Hence, $||\dot{\nabla}^2\xi||_{2,-2}\le||\dot{\nabla}^2\xi||_{2,-5/2}$ for smooth $\xi$.\\ 
Since $C_c^{\infty}$ is dense in $W_{-1/2}^{2,2}$, it follows that above estimate holds  
for all $\xi\in W_{-1/2}^{2,2}$.\\ To get the final estimate, we need to use first weighted Poincare inequality and then Sobolev inequality :\\ 
Thus, we have $||u||_{p,\delta}\le c ||\dot{\nabla}u||_{p,\delta-1}\le c||\dot{\nabla}^2u||_{p,\delta-2}$. Moreover, \\
\begin{equation*}
\begin{split}
||u||_{2,2,-1/2}=\sum\limits_{j=0}^2||\dot{\nabla}^2u||_{2,\delta-j}& =||u||_{2,-1/2}+||\dot{\nabla}u||_{2,-3/2} \\
& +||\dot{\nabla}^2u||_{2,-5/2}\le c||\dot{\nabla}^2u||_{2,-5/2}
\end{split}
\end{equation*}
Therefore, $||\xi||_{2,2,-1/2}\le c ||\dot{\nabla}^2\xi||_{2,-5/2}$ \\
Also, $||\pi||_{6,-3/2}\le c ||\pi||_{1,2,-3/2}$  and  $||X||_{6,-3/2}\le c ||X||_{1,2,-3/2}$. This follows from Sobolev inequality . \\
Combining different estimates derived above, and arranging them properly, we get the final result as follows,\\
$$||\xi||_{2,2,-1/2}\le c(||f_1||_{1,2,-3/2}+|| f_2, f_3||_{2,-5/2} + ||\xi||_{1,2,0})$$\\
Lemma 3.4 is thus proved.\\

To proceed further, we restructure $D\Phi^{\ast}$ into the operator $P^{\ast}$ defined by,
\begin{equation*}
\begin{split}
P^{\ast}(\xi)=\begin{pmatrix}
 & -g^{1/4}\nabla_p(2K_j^iN+L_Xg_j^i) \\
& -g^{1/4}(\gamma N/4\beta+L_X\psi_j^i)\\
&[g^{1/4}(\nabla^i\nabla_jN-\delta_j^i\triangle_gN+(S_j^i-E_j^i)N)+2\beta N\psi_i\psi_j \\
&-\beta Ng\psi_i\psi^i-\beta N\gamma^2g-g^{-1/4}L_X\pi_j^i] \\
&-(4\beta N\triangle\psi+4\beta \nabla N.\nabla\psi)g^{1/4}+g^{1/4}L_X\gamma_j^i 
\end{pmatrix}
\end{split}
\end{equation*}
We now prove the following Lemma:\\
\textbf{Lemma 3.5:} $P^{\ast}:W_{-1/2}^{2,2}(T)\to L_{-5/2}^2$ is bounded and satisfies: \\
$||\xi||_{2,2,-1/2}\le c||P^{\ast}\xi||_{2,-5/2} + C||\xi||_{1,2,0}$  where $C$ depends on $||(g,\psi,\pi,\gamma)||_\mathcal{P}$‖ \\
and $P^{\ast}=P^{\ast}_{(g,\psi,\pi,\gamma)}$ has Lipschitz dependence on $(g,\psi,\pi,\gamma)\in \mathcal{P}$, i.e. 
$$||(P_{(g,\psi,\pi,\gamma)}^{\ast}-P^{\ast}_{(\tilde{g},\tilde{\psi},\tilde{\pi},\tilde{\gamma})}\xi||_{2,-5/2}$$
$$\le {C_1}||(g-\tilde{g},\psi-\tilde{\psi},\pi-\tilde{\pi},\gamma-\tilde{\gamma})||_\mathcal{P}||\xi||_{2,2,-1/2}.$$
where ${C_1}$ depends on $||(g,\psi,\pi,\gamma)||_\mathcal{P}$  and $||(\tilde{g},\tilde{\psi},\tilde{\pi},\tilde{\gamma})||_\mathcal{P}$ \\
\textbf{Proof:} We have $P^{\ast}$ is bounded, that is $||P^{\ast}_{(g,\psi,\pi,\gamma)}\xi||_{2,-3/2} \le C ||\xi||_{2,2,-1/2}$\\ 
follows from the estimates analogous to those of Lemma 3.4. The elliptic estimate :   $||\xi||_{2,2,-1/2}\le c||P^{\ast}\xi||_{2,-5/2}+ c||\xi||_{1,2,0}$ \\
directly follows from $$||\xi||_{2,2,-1/2}\le c(||f_1||_{1,2,-3/2}+||f_2, f_3||_{2,-5/2} +||\xi||_{1,2,0})\ (\text{from Lemma 3.4})$$

As regards Lipschitz dependence, we find estimates for $(P^{\ast}_{(g,\psi,\pi,\gamma)}-P^{\ast}_{(\tilde{g},\tilde{\psi},\tilde{\pi},\tilde{\gamma})})\xi$ by considering its individual componenets. \\
To begin with, we note that $||g-\tilde{g}||_{\infty}, ||(N,X)||_{\infty}$ are bounded by $||g-\tilde{g}||_{2,2,-1/2},||\xi||_{2,2,-1/2}$ respectively. \\
Proceeding as in Bartnik [$2$], since $\nabla-\tilde{\nabla}\equiv \dot{\nabla}(g-\tilde{g})$, by using (2) \\ 
we obtain 
$$||(\nabla-\tilde{\nabla})D\Phi_g^{\ast}\xi||_{2,-5/2}\le c||\dot{\nabla}(g-\tilde{g})||_{1,2,-3/2}||D\Phi_g^{\ast}\xi||_{1,2,-3/2}$$
Also, $D\Phi_g^{\ast}\xi=-2 (NK_{ij})+\nabla (_i X_j)$. This gives 
$$||D\Phi (g,\psi,\pi,\gamma)_g^{\ast}\xi - D\Phi (\tilde{g},\tilde{\psi},\tilde{\pi},\tilde{\gamma})_g^{\ast}\xi||_{1,2,-3/2}$$
$$\le c ||N (K- \tilde{k})||_{1,2,-3/2}+c||\dot{\nabla} (g-\tilde{g})X||_{1,2,-3/2}\hspace{1.5cm}   (4) $$     
The first term on the right hand side above is estimated by \\
$||N||_{\infty}||K-\tilde{k}||_{1,2,-3/2}+||\dot{\nabla}N (K-\tilde{k})||_{2,-5/2}$, and similarly for the second term .\\ 
Again using the $L^{\infty}$ bound and equation $(2)$, the above difference in equation (4) is controlled by $c ||\xi||_{2,2,-1/2}$. \\
Now consider $D\Phi_\psi^{\ast}\xi= \gamma N/4\beta+L_X\psi=\gamma N/4\beta+X\nabla\psi$\\
For this term, we have \\ 
\begin{equation*}
\begin{split}
& ||D\Phi (g,\psi,\pi,\gamma)_\psi^{\ast}\xi-D\Phi (\tilde{g},\tilde{\psi},\tilde{\pi},\tilde{\gamma})_\psi^{\ast}\xi||_{1,2,-3/2}\\
& \le C||N(\gamma-\tilde{\gamma})||_{1,2,-3/2}+C||\nabla (\psi-\tilde{\psi})X||_{1,2,-3/2} \hspace{1.5cm} (5)
 \end{split}
\end{equation*}
As before, first term of (5) is $\le $\\
$\||N||_{\infty}||(\gamma-\tilde{\gamma})||_{1,2,-3/2}+||\dot{\nabla}N(\gamma-\tilde{\gamma})||_{2,-5/2}$  and simillarly for the second term.\\
Again difference in equation $(5)$ is controlled by $C||\xi||_{2,2,-1/2}$. \\ 
Similar estimates can be found for $D\Phi(g,\psi,\pi,\gamma)_\pi^{\ast}\xi - D\Phi (\tilde{g},\tilde{\psi},\tilde{\pi},\tilde{\gamma})_\pi^{\ast}\xi$ and $D\Phi(g,\psi,\pi,\gamma)_\gamma^{\ast}\xi-D\Phi (\tilde{g},\tilde{\psi},\tilde{\pi},\tilde{\gamma})_\gamma^{\ast}\xi$.\\
\begin{equation*}
\begin{split}
\text{Therefore,}\ & ||P_{(g,\psi,\pi,\gamma)}^{\ast}-P^{\ast}(\tilde{g},\tilde{\psi},\tilde{\pi},\tilde{\gamma}))\xi||_{2,-5/2} \\
& \le {C_1}||(g-\tilde{g},\psi-\tilde{\psi},\pi-\tilde{\pi},\gamma-\tilde{\gamma})||_\mathcal{P}||\xi||_{2,2,-1/2} \\
\end{split}
\end{equation*}\\
where ${C_1}$ is an appropriate constant.\\
Thus Lemma 3.5 is prooved.\\
Next step is to show that weak solutions of the equation \\
\begin{equation*}
\begin{split}
 D\Phi^{\ast}_{(g,\psi,\pi,\gamma)}(\xi)=(f_1 ,f_2,f_3,f_4)\\
\end{split}
\end{equation*}\\
satisfy the elliptic estimate. The procedure to prove this result for coupled Einstein and scalar fields is exactly the same as in Bartnik [$2$], and we assume the following lemma as proved there :\\

\textbf{Lemma 3.6:} If $\xi = (N,X^i)$ is a weak solution of $D\Phi^{\ast}_{(g,\psi,\pi,\gamma)}(\xi)=(f_1 ,f_2,f_3,f_4)$ with $(f_1,f_2,f_3)\in W_{-3/2}^{1,2}\times L_{-5/2}^2\times L_{-5/2}^2$ and $(g,\psi,\pi,\gamma)\in \mathcal{P}$, then $\xi\in W_{-1/2}^{2,2}$ is a strong solution and satisfies estimate of Lemma 3.4. \\\\
Our next step is to prove that the kernel of $D\Phi^{\ast}$ is trivial in the space of lapse-shift functions which decay at infinity. This will make the operator $D\Phi^{\ast}$ injective. Later, in Section $4$, we prove, as a part of our main theorem, that the operator $D\Phi$ has a closed range. Combining these two results, and applying Fredholm alternative, we finally conclude that $D\Phi$ is surjective. Hence the Implicit Function Theorem is applicable giving the desired Hilbert manifold structure of the solution set $\Phi^{-1}(0,0)$.\\
 As a consequence of a series of results, we arrive at the following theorem :\\
\textbf{Theorem 3.7:} Suppose $\Omega \subset M$ is a connected domain and $E_R \subset\Omega$ for some exterior domain $E_R$. Let $(g,\psi,\pi,\gamma)\in \mathcal{P}$ and suppose $\xi$ satisfies $D\Phi (g,\psi,\pi,\gamma)^{\ast}\xi=0$ in $\Omega$. Then $\xi\equiv 0$ in $\Omega$. \\
The results required to prove this theorem, and the proof of the theorem itself follow exactly as in Bartnik [$2$] in our case, and we omit all these proofs. For details, we refer the reader to Bartnik [$2$].\\

Thus kernel of $D\Phi^{\ast}$ is trivial and as discussed above, it remains to show that the operator $D\Phi$ has a closed range. This is proved in the following section. Applying the Implicit Function Theorem, we then conclude that $C$ is a smooth Hilbert submanifold of $\mathcal{P}$.\\\\
\item \textbf{Proof of Main Theorem :}\\
\textbf{Theorem 3.1:}  For each $(\in,S_i)\in \mathcal{L}^{\ast}$, the constraint set, \\
$C (\in, S_i)=\left\{(g,\psi,\pi,\gamma)\in \mathcal{P}: \Phi(g,\psi,\pi,\gamma)=(\in,S_i)\right\}$ is a Hilbert submanifold of $\mathcal{P}$.\\ 
In particular, the space of solutions of the constraint equations for coupled Einstein and scalar fields, $C=\Phi^{-1}(0)=C(0,0)$ is a Hilbert manifold. \\
To prove this, we use previous lemmas and the Implicit Function Theorem. \\ 
\textbf{Proof:} To apply the Implicit Function Theorem, we must show that $D\Phi : \mathcal{G}^+\times \mathcal{F}_{s,\delta}\times \mathcal{K}\times\hat{\mathcal{F}}_{s-1,\delta+1}\rightarrow \mathcal{L}^{\ast}$ is surjective and splits. Since  $D\Phi$ is bounded, its kernel is closed and hence splits. We have shown in above theorem that,\\
$Ker\left\{ D\Phi (g,\psi,\pi,\gamma)^{\ast}\right\}= \left\{0\right\}$, so the cokernel of $D\Phi$ is trivial.\\
Thus to show that $D\Phi$ is surjective, it is sufficient to show that it has closed range. We prove this by direct argument.\\
We consider particular variations $(\tilde{g},\tilde{\psi},\tilde{\pi},\tilde{\gamma})$ of $(g,\psi,\pi,\gamma)$ determined from fields $(y,Y^i)$ of the form, $\tilde{g}_{ij}=2yg_{ij}$,\\
$\tilde{\pi}^{ij}=(\nabla^iY^j+\nabla^jY^i-\nabla_kY^kg_{ij})\sqrt{g}$.\\
We restrict $D\Phi$ to particular variations, namely, $(\tilde{g},\tilde{\psi},\tilde{\pi},\tilde{\gamma})\in  T_{(g,\psi,\pi,\gamma)}\mathcal{P}$, such that $D\Phi$ resembles an elliptic operator. In particular, we write $(\tilde{g},\tilde{\psi},\tilde{\pi},\tilde{\gamma} )= f(y,Y)$ and thus $D\Phi(\tilde{g},\tilde{\psi},\tilde{\pi},\tilde{\gamma}) = D\Phi[f(y,Y)]= F(y,Y)$
Considering this restricted tangent space, we define $F(y,Y)=D\Phi(g,\psi,\pi,\gamma).(\tilde{g},\tilde{\psi},\tilde{\pi},\tilde{\gamma})$ 
\begin{equation*}
\begin{split}
=\begin{pmatrix}
& \{\mu(g)^{-1}\left\{-\frac{1}{2}\left(\pi.\pi-\frac{1}{2}(tr\pi)^2tr\tilde{g}\right)+2\left(\pi-\frac{1}{2}(tr\pi)g\right)\tilde{\pi}+2\left(\pi\times\pi-\frac{1}{2}(tr\pi)\pi\right)\tilde{g}\right\}\\
&-\mu(g)[\delta\delta\tilde{g}+\triangle (tr\tilde{g})-Ein(g)\tilde{g}]-\beta\gamma^2g\tilde{g}\mu(g)-\beta(2\tilde{\psi}-g|\nabla\psi|^2\tilde{g}\mu (g) \\
& 4\beta\gamma\eta\mu(g)+\nabla\psi\nabla\tilde{\psi}\mu(g)\}-2\tilde{g}\delta g\pi-2g\delta g\tilde{\pi}+\pi^{jl}(\tilde{g}_{kj;l}+\tilde{g}_{kl;j}-\tilde{g}_{jl;k})-\gamma\nabla\tilde{\psi}-\tilde{\gamma}\nabla \psi
\end{pmatrix}
\end{split}
\end{equation*}
We now require the scale broken estimate for operators which are asymptotic to the Laplacian. Towards this we use the following propositions.\\
\textbf{Proposition 4.1:}  If $u\in L_{\delta}^p$ and $Pu\in L_{\delta-2}^2$, with $1<p\le q$ and $\delta\in R$, then $u\in W_{\delta}^{2,p}$ and satisfies, $||u||_{2,p,\delta}\le C(||Pu||_{p,\delta-2}+||u||_{p:BR})$ where $R$ is fixed and is independent of $u$. \\
This proposition is proved in [3].\\
\textbf{Proposition 4.2:} The map $f: W_{-1/2}^{2,2}\rightarrow T_{(g,\psi,\pi,\gamma)}\mathcal{P}$, and therefore also the map, $F:W_{-1/2}^{2,2}\rightarrow L_{-5/2}^2$, is a bounded operator. \\ 
Proof of this follows similar to the proof of lemma (4.14) in McCormik [15]and we omit it.\\
Using proposition (4.1) we establish a scale-broken estimate for $F$, which will complete the proof of the theorem.
That is, we have to prove that for $\mathcal{Y}=(y,Y)\in W_{-1/2}^{2,2},F$ satisfies  the estimate: 
$$||\mathcal{Y}||_{2,2,-1/2}\le C (||F [\mathcal{Y}]||_{2,-5/2}+||\mathcal{Y}||_{2,0})$$ 
Using weighted Holder and Sobolev inequalities we have,
$$||\nabla^2\tilde{g}||_{2,-5/2}\le C||y||_{2,2,-1/2}$$
$$||\nabla^2\tilde{\pi}||_{2,-5/2}\le C ||Y||_{2,2,-1/2}$$
And then by weighted Poincare inequality, we have that $f$ is bounded. It follows immediately that $F$ is also bounded. Also, we have the general scale-broken estimate for $\triangle (cf.[$3$])$ : 
$$||u||_{2,2,-1/2}\le C(||\triangle u||_{2,-5/2}+||u||_{2,0})$$
The scale broken estimate for $F$ is then obtained by comparing $F$ to the Laplacian. We write $F[\mathcal{Y}]= (F_1, F_2)$ for the sake of presentation, and bound the terms separately as follows: \\
The norms of $\pi,\gamma,\nabla g$ are all finite and can be merged into constant $C$. \\
By using, weighted inequalities, Young’s inequality and the definition of the $W_{\delta}^{k,p}$ norm directly, we have further estimates,
\begin{equation*}
\begin{split}
||\xi||_{\infty,0}& \le C||\xi||_{1,4,0}=||\xi^{1/4}\xi^{3/4}||_{1,4,0} \\
& \le C||\xi^{1/4}||_{1,8,0}||\xi^{3/4}||_{1,8,0}\le C||\xi||_{1,2,0}^{1/4}||\xi||_{1,6,0}^{3/4}\le C||\xi||_{1,2,0}^{1/4}||\xi||_{2,2,0}^{3/4} \\
& \le C||\xi||_{1,2,0}^{1/4}\left(||\xi||_{1,2,0}+||\nabla^2\xi||_{2,-2}\right)^{3/4} \\
& \le C||\xi||_{1,2,0}+C(||\xi||_{1,2,0}+||\nabla^2\xi||_{2,-2}) \\
& \le C ||\xi||_{1,2,0}+||\nabla^2\xi||_{2,-2} 
\end{split}
\end{equation*}
\begin{equation*}
\begin{split}
\text{Also},\ ||\nabla\xi||_{3,-1} & \le C||\xi||_{1,3,0}=||\xi^{1/3}\xi^{2/3}||_{1,3,0}\le C||\xi^{1/3}||_{1,6,0}||\xi^{2/3}||_{1,6,0} \\
&\le C||\xi||_{1,2,0}^{1/3}||\xi||_{1,4,0}^{2/3}\le C||\xi||_{1,2,0}^{1/3}||\xi||_{2,2,0}^{2/3} \\
&\le C||\xi||_{1,2,0}^{1/3}(||\xi||_{1,2,0}+||\nabla^2\xi||_{2,-2})^{2/3} \\
& \le C||\xi||_{1,2,0}+\in (||\xi||_{1,2,0}+||\nabla^2\xi||_{2,-2})\\
&\le C||\xi||_{1,2,0}+||\nabla^2\xi||_{2,-2}
\end{split}
\end{equation*}
Also using Holder , Sobolev and Interpolation inequalities, we have a relation,
\begin{equation*}
\begin{split}
& ||\pi.\nabla u||_{2,-5/2}^2\le C||\nabla u||_{3,-1}||\pi||_{6,-3/2}\le C||\pi||_{1,2,-3/2} ||\nabla u||_{3,-1}\\
& \le C||u||_{2,2,-1/2}+C||u||_{2,0}
\end{split}
\end{equation*}
$||\nabla^2\psi||$ is also finite and satisfies above inequalities. \\
Therefore we can write,
$$||\triangle y||_{2,-5/2}\le C||F_1||_{2,-5/2}+C||\mathcal{Y}||_{1,2,0}+\in ||\nabla^2\mathcal{Y}||_{2,-2} \hspace{1.5cm} (7) $$
Similarly,
$$||\triangle Y||_{2,-5/2}\le C||F_2||_{2,-5/2}+(||\mathcal{Y}||_{\infty,0}+||\nabla \mathcal{Y}||_{3,-1})\hspace{1.8cm} (8) $$
Combining equations (7) and (8) we can write, 
$$||\triangle \mathcal{Y}||_{2,-5/2}\le C||F[\mathcal{Y}]||_{2,-5/2}+C(\in) ||\mathcal{Y}||_{1,2,0}+\in ||\nabla^2\mathcal{Y}||_{2,-2}$$
For $\in >0$, by inserting this into the scale-broken estimate for $\triangle$, we have,
$$||\mathcal{Y}||_{2,2,-1/2}\le C||F [\mathcal{Y}]||_{2,-5/2}+C (\in)||\mathcal{Y}||_{1,2,0}+\in ||\nabla^2\mathcal{Y}||_{2,-2}$$ 
The weighted interpolation and Poincare inequalities then give, 
$$||\mathcal{Y}||_{2,2,-1/2}\le C ||F[\mathcal{Y}]||_{2, -5/2 }+ C (\in)||\mathcal{Y}||_{2,0} + \in ||\nabla^2\mathcal{Y}||_{2,-2}$$ 
Finally, choosing $\in$ sufficiently small, we arrive at the scale broken estimate for $F$: 
$$||\mathcal{Y}||_{2,2,-1/2}\le C (||F[\mathcal{Y}]||_{2,-5/2} +C||\mathcal{Y}||_{2,0} )\hspace{2.5cm} (9) $$
Now the adjoint $F^{\ast}$ has a similar structure and the same argument shows that $F^{\ast}$ also satisfies an estimate $(9)$. By ellipticity estimate for $F^{\ast}$, it follows that $F^{\ast}$ has finite dimensional kernel. Hence, $F$ has closed range (from(9))with finite dimensional cokernel. Since clearly, range $F \subset range D\Phi$, we have shown that $D\Phi$ has closed range and the proof of theorem (3.1) is complete.\\
\item \textbf{Concluding Remarks :}\\
In this paper, we have proved that the set of solutions of constraint equations for coupled Einstein and scalar fields in classical general relativity possesses Hilbert manifold structure. This is proved in the context of asymptotically flat space-times. Similar results for Einstein-Yang-Mills system have been proved recently by McCormick [15,16] as a part of his Ph.D. thesis. If spacetime admits a compact Cauchy hypersurface, then smooth manifold structure of set of solutions of constraint equations corresponds to the absence of Killing fields for the spacetime. This is equivalent to saying that $Ker\left\{ D\Phi ^{\ast}\right\}$ is trivial. In general, if this kernel is not trivial, then under constant mean curvature condition on Cauchy hypersurface, $D\Phi^{\ast}$  comes out to be an elliptic operator and hence has finite dimensional kernel. This kernel is isomorphic to the space of Killing fields that the spacetime admits. In this situation, structure theory is needed. (See, for example, [$1,11,20$]). For asymptotically flat spacetimes, such structure theory is not needed even if spacetime admits Killing fields. This has been explained in Bartnik [$2$].\\
 Existence and uniqueness of solutions of constraint equations under different conditions on the mean curvature of spacelike (Cauchy) hypersurface is another important problem which has attracted attention of leading researchers since past two decades or more. We refer the reader to Isenberg [14] for latest review on this problem. Of late, using results from Bifurcation theory, Holst and Meier [13] proved that for constant mean curvature (CMC) hypersurfaces as well as for non-CMC hypersurfaces, solutions of constraint equations are not unique. By combining the work of Choquet-Bruhat, Isenberg and Pollack [6] with the techniques of [13], we wish to study non-uniqueness problem for solutions of constraint equations for coupled Einstein and scalar fields. This will be the topic of our future work.
\end{enumerate}
\textbf{References}
\begin{enumerate}[{[1]}]
\item Arms J., Marsden J.E.  and Moncrief V., The structure of the space of solutions of Einstein’s equationsII. Several Kiling fields and the Einstein-Yang-Mills Equations, Annals of Physics, Vol. 144 (1) (1982), 81-106
\item Bartnik R., Phase space for the Einstein equations, arXiv:gr-qc/0402070v1
\item Bartnik R., The mass of an asymptotically flat manifold, Communications on Pure and Applied Mathematics, Vol. XXXIX, (1986), 661-693.
\item Bartnik R. and Isenberg J., “The constraint equations”, in The Einstein equations and the large scalebehavior of gravitational fields (Eds. ChruscielP.T. and H. Friedrich H. ), Birkh¨auser, Basel, (2004),1–39, (arXiv:gr-qc/0405092v1)
\item Choquet-Bruhat Y., “General Relativity and The Einstein Field Equations”, Oxford monographs in Physics, Clarendon Press, Oxford, 2009
\item Choquet-Bruhat Y., Isenberg J. and Pollack D., The constraint equations for the Einstein-scalar field system on compact manifolds. Classical Quantum Gravity, 24(4),(2007), 809–828. 
\item Choquet-Bruhat Y. and York J.W., The Cauchy Problem. In: Held A. (Ed.) General Relativity and Gravitation, I. Plenum Press, New York,(1980), 99-172.
\item Christodoulou D. and Choquet-Bruhat Y., Elliptic systems in Hilbert spaces on manifolds which are euclidean at infinity. Acta Math. Vol.146, (1981), 129 - 150
\item Chrusciel P.T. and Delay E., Manifold structures for sets of solutions of the general relativistic constraint equations,  arXiv:gr-qc/0309001v2
\item Fischer A. and Marsden J.E., Topics in the dynamics of general relativity, in “Isolated gravitating systems in general relativity”, Ed. J. Ehlers, Italian Physical Society (1979), 322-395
\item Fischer A.,Marsden J.E. and Moncrief V., The structure of the space of solutions of Einstein’s equations.I. One Kiling field,Annales de la Institut H. Poincare, Section A, Vol.33 (2) (1980), 147-194
\item Hawking S.W. and Ellis G.F.R., large scale structure of spacetime , Cambridge university press , Cambridge , 1973.
\item Holst M. and Meier C., Non-uniqueness of solutions to the conformal formulation,  arXiv:1210.2156v2 [gr-qc] 
\item Isenberg J., “The initial value problem in general relativity”, chapter in "The Springer Handbook of Spacetime," edited by A. Ashtekar and V. Petkov. (Springer-Verlag), (2014), 303-321 (arXiv:1304.1960v1 [gr-qc]) 
\item McCormick S., The Phase Space for the Einstein-Yang-Mills Equations and the First Law of Black Hole Thermodynamics, Adv. Theor. Math. Phys. 18(4), (2014), 799-820
\item McCormick S., The Phase Space for theEinstein-Yang-Mills Equations,Black Hole Mechanics, and aCondition for Stationarity, Ph.D. thesis submitted to Monash University, 2014
\item Misner C., Thorne K. and Wheeler J., “Gravitation”, Chapter 21, Freeman Press, Sanfransisco, 1972.
\item Saraykar R.V. and Joshi N.E., Linearisation stability of Einstein’s equations coupled with self-gravitating scalar fields, Jour. Math. Phys. Vol.22 no.2, (1981), 343-347; Erratum Vol.23, (1982), 1738.
\item Saraykar R.V., Linearisation stability of coupled gravitational and scalar fields in Asymptotically flat space-times, Pramana, Vol.19, No.1,  (1982), 31-41.
\item Saraykar R.V., The structure of the space of solutions of Einstein’s equations coupled with Scalar fields, Pramana. Vol.20, No.4, (1983), 293-303.
\item York J.W., “Kinematics and dynamics of general relativity”, in “Sources of gravitational radiation”, Ed. L. Smarr, Cambridge (1979), 83-126

\end{enumerate}

\end{document}